\documentclass[preprint]{iucr}

\papertype{FA}
\paperlang{english}

\usepackage{graphicx}
\usepackage{upgreek}

\newcommand{\figsize}{0.5\columnwidth}

\begin{document}

\title{Detecting Orientational Order in Model Systems by X-ray Cross Correlation Methods}

\cauthor[a,b]{Felix}{Lehmk\"uhler}{felix.lehmkuehler@desy.de}{}
\author[a,b]{Gerhard}{Gr\"ubel}
\author[a,b,c]{Christian}{Gutt}

\aff[a]{Deutsches Elektronen Synchrotron (DESY), Notkestr. 85, 22607 Hamburg, \country{Germany}} 
\aff[b]{The Hamburg Centre for Ultrafast Imaging, Luruper Chaussee 149, 22761 Hamburg, \country{Germany}}
\aff[c]{current address: Department Physik, Universit\"at Siegen, Walter-Flex-Str. 3, 57072 Siegen, \country{Germany}}
\shortauthor{Lehmk\"uhler, Gr\"ubel and Gutt}

\maketitle

\begin{synopsis}
Computational X-ray cross correlation results from two-dimensional model strutures are presented and demonstrate how to extract the orientational local order of disorded samples. Special attention is spent on proper ensemble averaging and experimental influences such as detector noise and wave front distortions.
\end{synopsis}

\begin{abstract}
We present the results of a computational X-ray cross correlation analysis (XCCA) study on two dimensional polygonal model structures. We show how to detect and identify the orientational order of such systems, demonstrate how to eliminate the influence of the "computational box" on the XCCA results and develop new correlation functions that reflect the sample's orientational order only. For this purpose, we study the dependence of the correlation functions on the number of polygonal clusters and wave vector transfer $q$ for various types of polygons including mixtures of polygons and randomly placed particles. We define an order parameter that describes the orientational order within the sample. Finally, we determine the influence of detector noise and non-planar wavefronts on the XCCA data which both appear to affect the results significantly and have thus to be considered in real experiments.
\end{abstract}

\maketitle

\section{Introduction}
The study of angular correlation functions in diffraction patterns promises to detect and quantify the usually hidden bond orientational order in amorphous materials \cite{pnas,clark,gibson}. The use of angular correlation functions was originally proposed by Kam \cite{kam} for the structure determination of single particles in dilute solutions. The idea was revived recently first for simulations of diffraction patterns \cite{saldin_prb2010,saldin_actacryst,saldin_njp2010,elser,saldin_oe2011} and finally experimentally for nanorods \cite{saldin_njp2010,saldin_prl2011}, dumbbells \cite{chen,starodub} and structures with triangular symmetry \cite{pedrini}. In those studies, the diffraction pattern of a single particle is extracted from an ensemble of patterns from a dilute solution of randomly oriented identical particles via higher-order correlation functions. Phase retrieval algorithms finally allow the determination of the structure of a single particle.

By using coherent and ultrashort pulses of X-rays the inherent spatial and temporal averaging process of conventional (incoherent) X-ray scattering experiments can be avoided. Instead, snapshots of the instantaneous positions of all particles in the beam are reflected in the coherent diffraction patterns, the so called "speckle patterns". In order to uncover the hidden local symmetries in disordered matter such as liquids and glasses from the speckle patterns properly defined higher order correlation functions have to be devised and applied to the data\cite{pnas,wochner}. This method is called X-ray Cross Correlation Analysis (XCCA) and represents a promising tool for the study of disordered samples such as fast relaxing liquids at X-ray free-electron-laser facilities \cite{emma}. Here, X-ray intensities, pulse lengths and coherence properties are achieved \cite{gutt11} that in principle allow to measure speckle patterns even from molecular and atomic liquids which promises to reveal their bond-ordered structure.

In recent work \cite{altarelli,kurta} the potential of XCCA to yield information on the structure of single clusters has been challenged. It was claimed that the cross correlation analysis reflects the cross correlation of a single polygonal cluster only for very dilute systems whereas this is not the case for more realistic dense systems. Since many theoretical publications in XCCA are using simple 2D disordered structures\cite{altarelli,saldin_njp2010,kurta} it is important to address the question to what extent the number of particles is compromising the access to single particle properties. 

In this paper we show how structural information can be extracted from speckle patterns of polygonal model structures by using properly defined correlation functions. For simplicity and in order to enable the comparison to previous theoretical studies, we are focusing here on two-dimensional systems to demonstrate and discuss the XCCA method. The discussion of 2D systems is of particular relevance to recent experimental studies, that made use of quasi two dimensional samples such as nanorods \cite{saldin_prl2011} or dumbbells \cite{chen}. We start with the discussion of coherent scattering from a single pentagon. Various simple simulations of model structures are used to demonstrate the performance of the correlation functions with respect to particle number, $q$-dependence, and snapshots of various configurations. Here, the two-dimensional models used so far are expanded to new correlation functions and more complex structures. Finally, we discuss the potential impact of the simulations on experimental studies.

\section{Scattering from two-dimensional samples}\label{sec:theo}
\subsection{Scattering from a single pentagon}
As a starting point we discuss the scattering from particles in a two dimensional pentagonal arrangement as a model for five-fold order. We assume particles arranged in a plane with fixed electron density forming the single pentagon. The incoming beam is orthogonal to the sample system. The scattering amplitude $A({\bf q})$ is described within the first Born approximation as the Fourier transform of the electron density $\rho({\bf r})$
\begin{equation}
A({\bf q})=\int \rho(\textbf{r}) \exp(i {\bf q}\cdot{\bf r})\mathrm{d}{\bf r},
\end{equation}
where \textbf{q} denotes the scattering vector. The electron density of the pentagon is given by
\begin{equation}\label{eq:el_dens}
\rho(r,\theta)=\frac{\delta(r-R_0)}{R_0}\sum_{j=1}^5\delta(\theta-\theta_j+\omega)
\end{equation}
in polar coordinates, with $R_0$ denoting the radius of the polygons, $\theta_j=\frac{2\pi}{5}\cdot j$, and $\omega$ denoting the absolute orientation of the pentagon with respect to a reference coordinate system. An expansion of the scattering amplitude in a Fourier series yields \cite{baddour}
\begin{equation}\label{eq:ftseries}
A(q,\theta)=\sum^{\infty}_{l=-\infty}F_l(q)\exp(il(\theta+\omega)),
\end{equation}
where $q=|{\bf q}|$.
The Fourier coefficients in Eq.~\ref{eq:ftseries} are
\begin{equation}
F_l(q)=i^{-l} J_l(qR_0)\sum_{j=1}^5 \exp(il\theta_j)
\label{eq:four_coeff}
\end{equation}
with $J_l$ denoting the Bessel functions of first kind. Because of the pentagonal symmetry, all coefficients with $(l\; \textrm{mod}\; 5) \neq 0$ in Eq.~\ref{eq:ftseries} are zero. Odd terms cancel out pairwise (e.g.~$l=5$ and $l=-5$) and only terms with $l$ being a integer multiple of 10 contribute to the scattering amplitude. Furthermore, because $F_l(q)\propto J_l(qR_0)$, higher order terms ($l>10$) become only observable at large $q$. The scattered intensity of such a cluster is given by $I(q,\theta)=A(q,\theta)A^*(q,\theta)$, thus we obtain the well known result that the local pentagonal structure is directly accessible from the symmetry of the scattering pattern. One way to determine this symmetry is the calculation of the angular power spectrum of $I(q,\theta)$, resulting in contributions of $l=10, 20, 30, \ldots$ depending on $q$ for the example discussed.

\subsection{Scattering from systems of randomly ordered polygons}
In our model study we require that the clusters are uncorrelated in space. Orientational disorder is fulfilled with the polygonal clusters being randomly rotated with respect to the axis of the incoming photon beam. The requirement for positional disorder is more difficult to realize when increasing the number of particles and when overlap of particle clusters shall be avoided. Here special care has to be taken in order to avoid introducing artificial correlations via the computer model.

\begin{figure}%
\centering
\includegraphics[width=0.7\columnwidth]{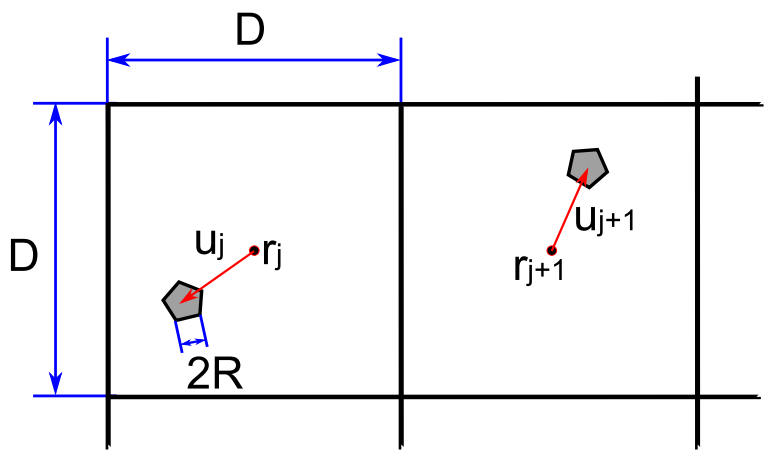}%
\caption{Representation of a tiled simulation model (model B). The particles are placed randomly at the vertices of polygons (particle distance $2R$) in squares with size $D\times D$. The vector ${\bf u}_j$ represents the random position with respect to the lattice point $r_j$. The sizes of the fields and pentagons shown here are consistent with the case of 1600 polygons ($D=7.5R$).}%
\label{fig:cell}%
\end{figure}

We have pursued two routes for the simulation. In model A we placed the clusters randomly in the computational box. This approach is working well for dilute systems but when increasing the number of particles overlap between clusters becomes more likely. In model B we used a tiling of the computational box with only one cluster allowed per tile (see Fig.~\ref{fig:cell}).
  
For both models the scattering amplitude can be calculated as follows. We start with a two dimensional lattice and allow arbitrary movements of the polygons around the lattice positions ${\bf r}_j$. The final distance to the initial lattice point is given by the displacement vectors ${\bf u}_j$. The two models basically differ in the size of the allowed displacement vectors ${\bf u}_j$. While model A accepts displacements as large as the computational box, model B limits the size of $u_j=|{\bf u}_j|$ so that the position of each cluster is limited to the tile.

The scattering amplitude becomes
\begin{equation}\label{eq:defB}
B({\bf q})=\sum\limits_j A_j({\bf q}) \exp(i{\bf q}({\bf r}_j+{\bf u}_j))=\sum\limits_j A_j({\bf q}) \exp(i{\bf q}{\bf r}'_j),
\end{equation}
with ${\bf r}'_j={\bf r}_j+{\bf u}_j$. This results in the scattering intensity
\begin{equation}
I({\bf q})=B({\bf q})B^*({\bf q})=\sum\limits_{j,k}A_j({\bf q})A_k^*({\bf q})\exp(i{\bf q} ({\bf r'}_j-{\bf r'}_k)).
\end{equation}
The sample's orientational correlations are revealed by performing an orientational Fourier analysis of the intensity. The corresponding Fourier coefficients are given by
\begin{equation}
I_l(q)=\sum\limits_{j,k} \widehat{A}_{j,k}(q,l) * \widehat{G}_{j,k}(q,l),
\label{eq:convol}
\end{equation}
with the orientational Fourier transforms of the clusters $\widehat{A}_{j,k}(q,l)=\mathrm{FT}(A_j({\bf q})A_k^*({\bf q}))$ and of the positional part $\widehat{G}_{j,k}(q,l)=\mathrm{FT}(\exp(i{\bf q} ({\bf r'}_j-{\bf r'}_k)))$. For liquids and glasses the ensemble averaged positional part is the well known structure factor $S(q)$ describing the typical short ranged positional correlations of particles. However, in the spirit of XCCA it can also contain information on higher order correlation functions and as such provide new insights into glasses and liquids.

Depending on the construction and properties of this positional correlator $\widehat{G}_{j,k}(q,l)$ it is immediately clear that it can contribute additional Fourier components to the XCCA signal. Diluted systems imply large values of the displacement vectors $u_j$ yielding small values of the correlator $\widehat{G}_{j,k}(q,l)$. When increasing the concentration $u_j$ usually gets smaller leading to an increase of $\widehat{G}_{j,k}(q,l)$ which becomes maximized for cluster positions fixed on lattice points. Moreover it is also important to note that the positional correlator scales as $N^2$ with $N$ being the number of particles. Thus even when $u_j$ is large, as in our model A, an increasing number of particles will lead to an increase of the size of the positional correlator. If $\widehat{G}_{j,k}(q,l)$ carries also angular Fourier components they will contribute to the XCCA signal depending on the size of the displacement vector $u_j$ and of the number of particles. Thus the effect of the positional correlator is of great importance when considering dense systems and we will show how it affects the XCCA signal and how to discriminate between the different XCCA signals in the analysis.

\section{X-ray Cross Correlation Analysis}
In order to detect a sample's orientational order, the power spectrum of the intensity is calculated with respect to $\varphi$ for every single speckle pattern \cite{altarelli} via
\begin{equation}
|I_l(q)|^2=\left|\int\limits_0^{2\pi}I(q,\varphi)\exp(il\varphi)\mathrm{d}\varphi\right|^2,
\label{eq:fti}
\end{equation}
where $I$ denotes the scattered intensity on an annulus with scattering vector $q$ and azimuthal position $\varphi$ (see Fig.~\ref{fig:sim} for definitions), $I_l(q)$ denotes the Fourier coefficients of $I(q,\varphi)$.

XCCA is based on measuring correlations within a single diffraction pattern. However, for sufficient statistical accuracy the correlation functions have to be calculated for many different diffraction patterns. Therefore we devise the correlation function
\begin{equation}
\Psi_l(q) \equiv \langle {I}_l(q)^2 \rangle_e - \langle {I}_l(q) \rangle_e^2.
\label{eq:psi}
\end{equation}
as variance of the Fourier coefficients $I_l(q)$ of the intensity $I(q,\phi)$. Here $\langle\cdot\rangle_e$ denotes an ensemble average from different scattering patterns with $\Psi_l(q)$ then being the variance over the ensemble of the Fourier coefficients $I_l(q)$. It is not affected by angular correlations from the positional correlator $\widehat{G}_{j,k}(q,l)$ because $\widehat{G}_{j,k}(q,l)$ is expected to be invariant to ensemble averaging and thus cancels out when calculating the variance in Eq.~\ref{eq:psi}. Hence, $\Psi_l$ reflects the single cluster's orientational order only.

In literature the cross correlation function
\begin{equation}
C(q,\Delta)=\frac{\langle I(q,\varphi)I(q,\varphi+\Delta) \rangle_\varphi - \langle I(q,\varphi) \rangle_\varphi^2}{\langle I(q,\varphi) \rangle_\varphi^2}.
\label{eq:cq1}
\end{equation}
is discussed frequently \cite{pnas,saldin_njp2010,saldin_prl2011,kurta_pre}. Here the average $\langle\cdot\rangle_\varphi$ denotes the angular average around rings of constant $q$. The sample's orientational order is accessed by the Fourier coefficients $C_l(q)$ of $C(q,\Delta)$. After replacing the intensity $I(q,\varphi)$ by a normalized quantity
\begin{equation}
I^N(q,\varphi)=\frac{I(q,\varphi)-\langle I(q,\varphi) \rangle_\varphi }{\langle I(q,\varphi) \rangle_\varphi},
\label{eq:inorm}
\end{equation}
the Fourier coefficients $C_l$ of $C(q,\Delta)$ equal the power spectrum of $I^N(q,\varphi)$ (Wiener-Khinchin theorem)
\begin{equation}
C_l(q)=|I^N_l(q)|^2,
\end{equation}
and thus ${C}_l(q)\propto|{I}_l(q)|^2$ for $l\neq 0$ and $\Psi_l(q) = \langle {C}_l(q) \rangle_e - \langle {I}^N_l(q) \rangle_e^2$

The search for the orientational order of the clusters using $C_l(q)$ requires $\langle {I}^N_l(q) \rangle_e=0$, which is difficult to achieve in simulations due to the natural tiling of the real space. Hence, contributions from the positional correlations cannot be neglected. These constraints in fact can be overcome by using $\Psi_l$ instead of $C$. For $\langle {I}^N_l(q) \rangle_e=0$, it is simple to deduce $\Psi_l(q)= \overline{C}_l \equiv \langle {C}_l(q)\rangle_e$. For the remaining part of the paper we will thus use $\Psi_l(q)$ for studying the orientational order.

It is important to mention, that in simulations and theoretical studies the bond order of amorphous samples is expressed by locally defined order parameters \cite{steinhardt,kawasaki,tanaka10}. Such order parameters reflect the orientational order of each particle's next-neighbour shell. In contrast, $\Psi_l$ and $\overline{C}_l$ are averaged over all particles in the diffraction volume (typically $>10^8$ particles) and are thus a measure of the mean orientational order of the sample.

\section{XCCA on model structures}
\begin{figure}%
\includegraphics[width=\columnwidth]{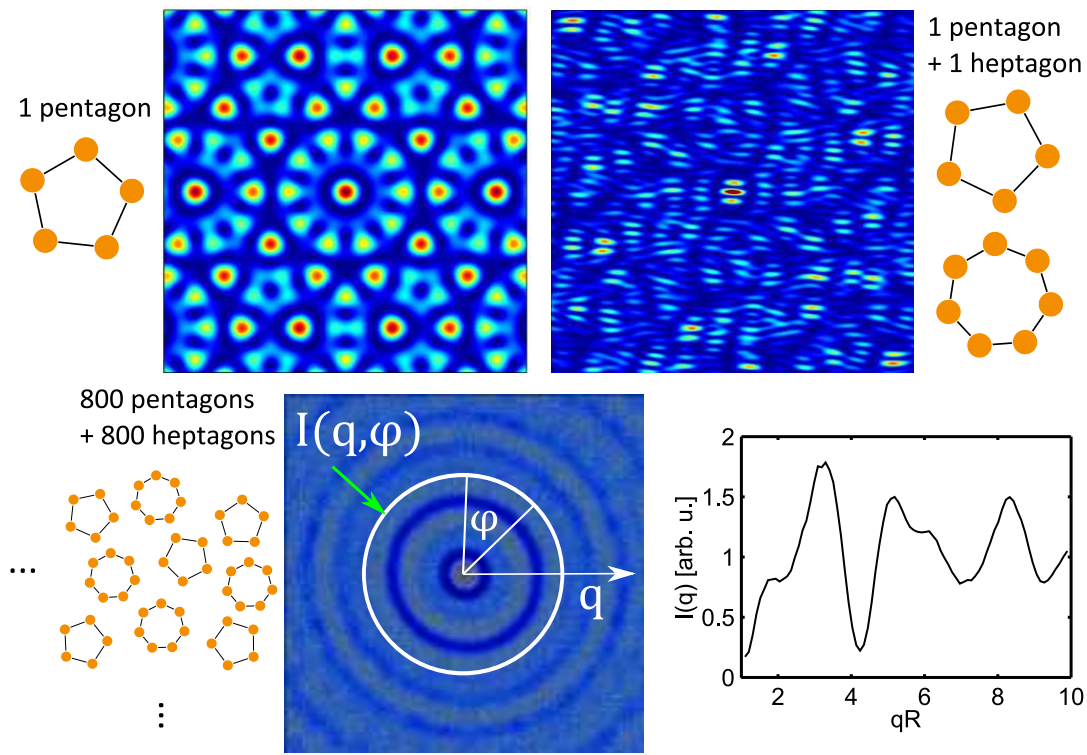}%
\caption{Calculated speckle patterns for one pentagon (up, left), one pentagon and one heptagon (up, right), 800 pentagons and 800 heptagons (bottom, left), and the integrated intensity $I(q)$ for 800 pentagons and 800 heptagons (bottom, right). In the bottom left figure, a ring of constant $q$ is shown which is used for the analysis.}%
\label{fig:sim}%
\end{figure}

Correlation functions were calculated for an ensemble of particles forming different polygonal structures that are placed into the two-dimensional computational box. To allow orientational disorder, each polygon is rotated by an arbitrary angle $\omega_j$ within the interval $[0, 2\pi/k]$ with $k$ denoting the number of vertices of the cluster. Fig.~\ref{fig:sim} shows typical speckle patterns of such polygon arrangements. For one pentagon, the symmetry is directly obvious from the pattern, while the patterns are smeared out with increasing number of particles. In this paper we focus on the analysis of orientational order in the next-neighbour distance which corresponds to the peak around $qR=\pi$ in the integrated intensity $I(q)$ shown in Fig.~\ref{fig:sim} (bottom right).

In the framework of XCCA, $\Psi_l(q)$ and $\overline{C}_l(q)$ were calculated from pattern ensembles of at least 1000 different cluster arrangements for the case of one pentagon and one heptagon and the case of 800 pentagons and heptagons, respectively. For 1600 polygons this corresponds to a volume fraction of 0.05 assuming spherical particles with radius $R$ at the edge of the pentagons. In this dilute limit we are able to focus on the orientational order of the polygonal structures neglecting cluster-cluster correlations. Within model A each cluster is placed to a random position into the computational box. 

\begin{figure}%
\centering
\includegraphics[width=\columnwidth]{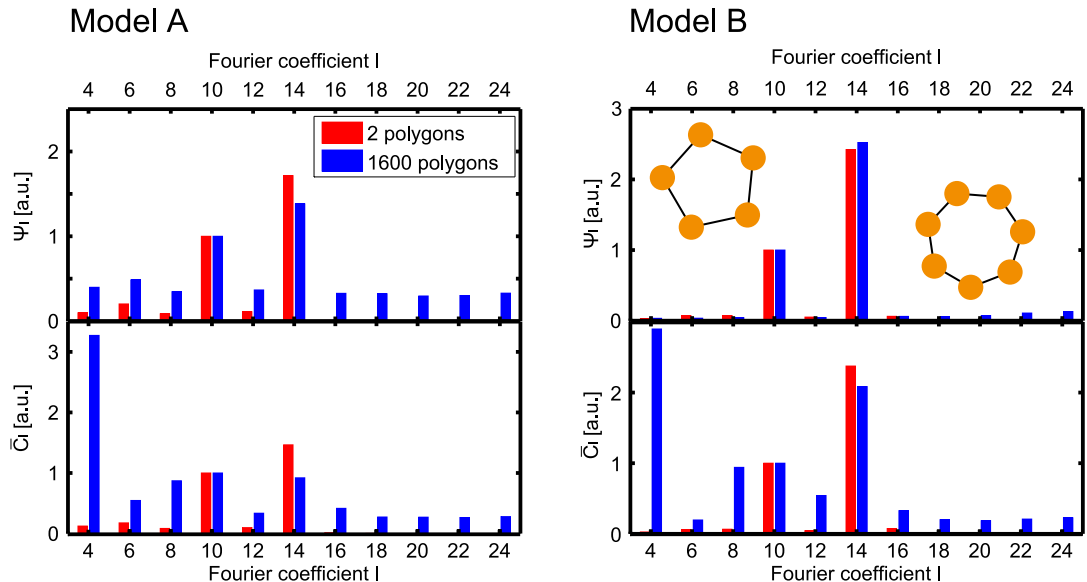}%
\caption{$\Psi_l$ (top), $\overline{C}_l$ (middle) and $\langle I_l \rangle$ (bottom) for the arrangement of 1 pentagon and 1 heptagon and 800 pentagons and 800 heptagons, respectively, at $qR=2\pi$ for model A and B as indicated. For comparison, the data are normalized to the value for $l=10$ (top and middle) and to $\sqrt{\Psi_{10}}$ (bottom), respectively.}%
\label{fig:pentahepta}%
\end{figure}

Model B uses a tiling of space to avoid overlapping. Therefor the computational box was divided in equally sized squares, in which the polygons are placed at random positions (i.e.~$|{\bf u}_j|\gg 2R$ in Eq.~\ref{eq:defB}, with $R$ denoting the particle radius, see Fig.~\ref{fig:cell}). Overlap is avoided as every square contains only one polygon. In this case, $\textbf{u}_j$ has to fulfill the boundary conditions
\begin{equation}
|\textbf{e}_{x/y}\cdot \textbf{u}_j|<\frac{D}{2}+R_P,
\end{equation}
where $\textbf{e}_{x/y}$ denotes the unit vectors in $x$- and $y$-direction, respectively, and $R_P$ the distance of the particle center to the center of the polygon. Within the tile random displacements of the clusters are allowed. In both models each polygon is rotated by a random angle $\omega_j$. From each speckle pattern the angular intensity distribution $I(q,\varphi)$ was calculated for the particular $q$-range of interest, i.e.~covering usually typical next-neighbour distances, followed by the calculation of both $\Psi_l(q)$ and $\overline{C}_l(q)$. The results are shown in Fig.~\ref{fig:pentahepta}. Odd symmetries do not contribute (see section \ref{sec:theo}), and the $l=2$ symmetry is dominated by Friedel's law \cite{saldin_njp2010,pnas}. Therefore, these components are not discussed. Moreover, only components up to $l=24$ are discussed for convenience, covering all relevant orientational orders of interest.

$\Psi_l$ shows for both polygon arrangements dominating contributions of $l=10$ and $l=14$. The maximum for $l=10$ reflects the five-fold symmetry of the pentagon, while the maximum for $l=14$ is a fingerprint of the heptagonal order. The different amplitudes originate from the different particle numbers in the polygons and scale roughly via $5^2/7^2$. For 1600 polygons other contributions become observable, in particular for model A. We attribute these contributions to the overlapping of polygon clusters, resulting in occurrence of further Fourier coefficients. This is supported by the results obtained within model B, where those contributions are much weaker. We conclude, that $\Psi_l(q)$ shows for both particle numbers only contributions that are connected to the polygonal symmetry. Additional components as seen e.g.~for model A seem to stem from overlapping clusters.

In contrast, as $\overline{C}_l$ is sensitive to constant contributions from the computational model such as e.g.~the tiling of space, Fourier coefficients are apparent that have no correspondance to the orientational order of the polygons. As such contributions are weak for a small number of polygons, we observe that $\overline{C}_l$ reflects the clusters' orientational order similar to $\Psi_l(q)$ in this case, see Fig.~\ref{fig:pentahepta}. For 1600 polygons the contributions from the tiling becomes dominant so that the orientational order can hardly be detected. In particular the strong contribution for $l=4$ indicates the effect of the underlying symmetry due to tiling.

\begin{figure}%
\includegraphics[width=\columnwidth]{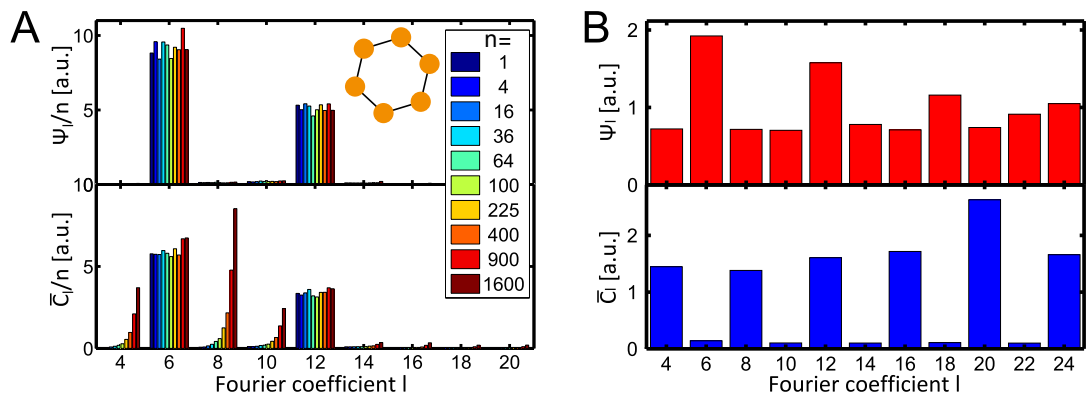}%
\caption{XCCA results for hexagon systems. (A) Influence of particle number on $\Psi_l$ and $C$. Data are normalized to the particle number $n$. Different colors represent different particle numbers as indicated. (B) $\Psi_l$ and $\overline{C}_l$ for 400 hexagons placed on a fixed square lattice.}%
\label{fig:hex_merged}%
\end{figure}

In order to investigate these effects in more detail, the influence of the particle number was studied for arrangements of 1 to 1600 hexagons. To avoid overlap of clusters we focus on model B in the following. Compared to pentagons or heptagons, hexagons exhibit a higher symmetry (hexagons show dominant $l=6$ and $l=12$ coefficients). The results are shown in Fig.~\ref{fig:hex_merged} A. For comparison, the results are normalized to the particle number $n$. As expected, $\overline{C}_l/n$ increases with $n$ reflecting the increasing dominance of the tiling expressed by $\overline{G}_{j,k}(q,l)$. Remarkably, the coefficients that reflect the orientational order of the polygonal clusters stay constant. $\Psi_l/n$ does on the other hand not exhibit any dependence on $n$ as observed before.

To demonstrate an extreme influence of the tiling, calculations of 400 hexagons on a fixed grid ($\textbf{u}_j=0$) only allowing for a random rotational orientation of each hexagon were performed. These confirm the shortcomings of $C$ in measuring orientational order and its sensitivity to an underlying computational grid, see Fig.~\ref{fig:hex_merged} B. $\overline{C}_l$ only reflects the underlying square order of the lattice placement of the hexagons, i.e.~strong maxima for $l=4,8,12,\ldots$. In contrast, $\Psi_l$ still peaks at the relevant Fourier coefficients that reflect the hexagonal order of the clusters.

\begin{figure}%
\centering
\includegraphics[width=\figsize]{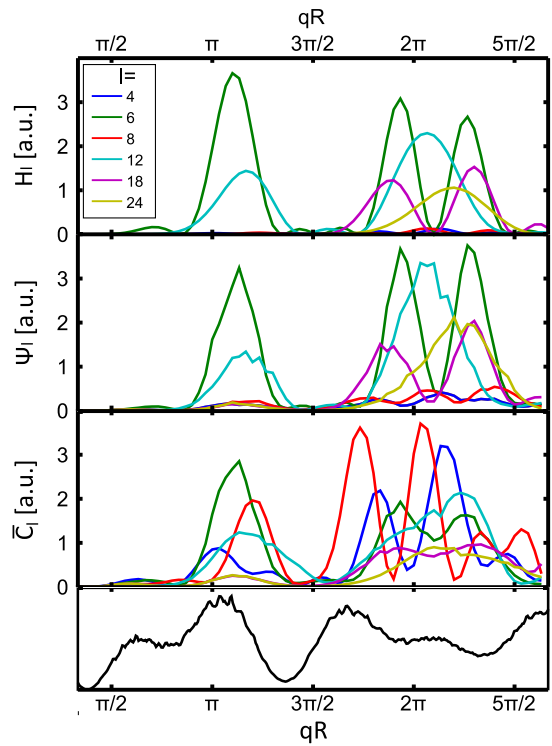}%
\caption{$q$-dependence of selected Fourier coefficients $l$ for arrangements of 400 hexagons. The corresponding values for $l$ are indicated. $H_l(q)$ represents the power spectrum of one single hexagon. The intensity $I(q)$ is shown at the bottom as black curve.}%
\label{fig:q}%
\end{figure}

In order to gain a deeper insight into the difference between $\overline{C}_l$ and $\Psi_l$, the $q$-dependence of the correlation functions was analyzed for arrangements of 400 hexagons, see Fig.~\ref{fig:q}. The black line at the bottom shows the intensity $I(q)$ of the arrangements of hexagons. In addition, the power spectrum of $I$ for a single hexagon $H_l(q)=|I_l(q)|^2$ is shown on top. The maxima and minima of $\Psi_l$ follow in general the shape of $I(q)$. Around $qR=\pi$, $\Psi_6$ and $\Psi_{12}$ dominate as expected for hexagons. In the range of $qR=2\pi$, higher harmonics $(l=18, 24, \ldots)$ occur as expected from the discussion in section \ref{sec:theo}. Altogether, $\Psi_l(q)$ resembles the hexagonal symmetry represented by $H_l(q)$ very well. In contrast, the $q$-dependence of $\overline{C}_l$ is dominated by the contributions originated by the computational model, shown exemplary for $l=4$ and $l=8$ in Fig.~\ref{fig:q}.

We conclude that the ensemble average of the cross correlation function $C$ is not an appropriate measure of orientational order of polygonal clusters in our model calculations. $\overline{C}_l$ and $\Psi_l$ exhibit the same information only if the ensemble averages of $I_l(q)$ vanish for all $l$ \cite{wochner} which is not valid for the simulations shown here.  While $\Psi_l(q)$ represents the sample's orientational order, $\overline{C}_l$ is in particular sensitive to the tiling of space expressed by the positional correlator. Furthermore, since $\Psi_l$ is calculated as a variance of intensities, parasitic scattering from slits which usually cannot be neglected in a coherent x-ray experiment does not affect $\Psi_l$ significantly, whereas the cross correlation function $C$ is strongly affected by such experimental constraints.

\section{Characterisation of increasing order}
\begin{figure}%
\includegraphics[width=\columnwidth]{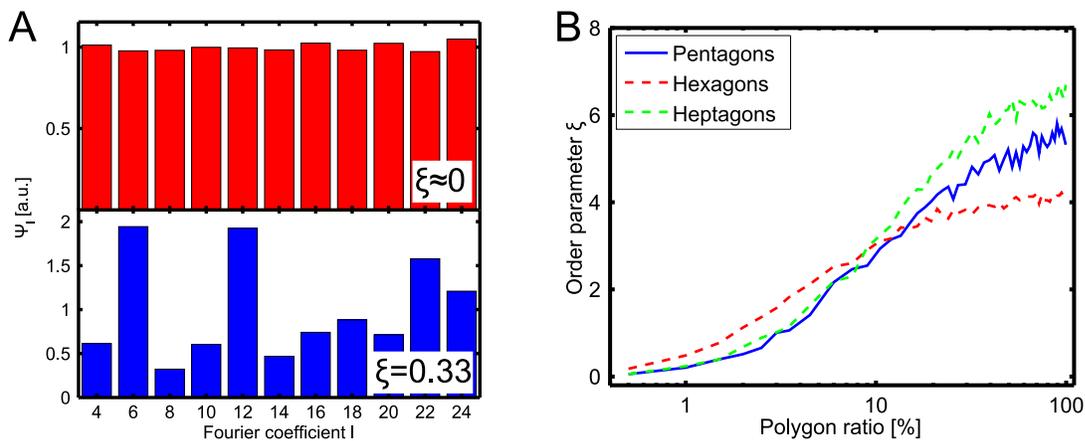}%
\caption{(A) Typical $\Psi_l$ for a system without dominating order of particle clusters (top) and a system showing dominating hexagonal order (bottom), here $l=6$ and $l=12$ dominate. $\Psi_l$ is normalized to its mean value $\langle\Psi_l\rangle_l$. (B) Order parameter $\xi$ for mixtures of polygons and randomly placed particles.}%
\label{fig:orderpsi}%
\end{figure}

In scattering experiments on amorphous samples, various Fourier coefficients reflecting the diversity of orientational order are expected. Typical Fourier coefficients for such samples are shown in Fig.~\ref{fig:orderpsi} A. Both samples are made out of 2400 single particles. The upper example represents $\Psi_l$ for a sample system that does not show any dominating order, while the bottom one shows dominating order, giving rise to a pronounced hexagonal order (maxima for $l=6, 12$). As discussed above, the absolute values of $\Psi_l/n$ reflect the strength of the corresponding Fourier coefficients $l$ and therefore represents an order parameter. However, in experimental situations the exact number of particles $n$ is usually unknown. Therefore, we define an order parameter that is sensitive to appearence of dominating order via
\begin{equation}
\xi(q)=\mathrm{var}\left( \frac{\Psi_l(q)}{\langle\Psi_l(q)\rangle_l}\right)_l,
\end{equation}
as the variance of a normalized $\Psi$ with respect to $l$. Here $\langle\Psi_l(q)\rangle_l$ denotes the mean value of $\Psi_l$ with respect to $l$, taking the all even terms between $l=4$ and $l=24$ into account. The examples in Fig.~\ref{fig:orderpsi} A exhibit $\xi_\mathrm{top}=6.1\cdot 10^{-4}\approx 0$ and $\xi_\mathrm{bottom}=0.33$, respectively. For the mixture of pentagons and heptagons (see Figs.~\ref{fig:sim} and \ref{fig:pentahepta}), we find $\xi_{5+7}=2.6$ at the typical next-neighbour distance, while a random arrangement of particles results in $\xi_\mathrm{rand}=7\cdot 10^{-5}\approx 0$. Therefore, $\xi$ is zero or close to zero in case of fully disordered samples, and increases with increasing order. Thus $\xi$ provides a measure of the occurrence of order in the sample.

The question emerges, how many ordered particles are necessary so that the order can be detected via calculation of $\xi$. Therefore we calculated $\xi$ for a mixture of polygons and randomly placed particles. The number of particles forming polygons $p$ range from 0\%, i.e.~a completely random system, up to 100\%\ consisting out of polygons only. Fig.~\ref{fig:orderpsi} B shows the result for such systems for pentagons, hexagons and heptagons, respectively. The $q$-range was chosen to fit the next-neighbour distance. In general, all systems show a similar shape with increasing $p$. Remarkably, a small number of ordered particles results already in a significant increase of $\xi$. For the pure polygon samples ($p=100$\%), we find $\xi_5=5.6$, $\xi_6=4.2$, and $\xi_7=6.7$ for pentagons, hexagons, and heptagons, respectively. The lower value for hexagons and the slightly different curve shape can be understood as caused by the occurrence of more Fourier components ($l=6, 12$) compared to the other polygons (e.g.~$l=10$ for pentagons) and thus a reduced variance $\Psi_l$. The random system ($p=0$\%) shows no dominant order, thus $\xi_\mathrm{rand}\approx 0$.

\section{Experimental constraints}
\begin{figure}%
\includegraphics[width=\columnwidth]{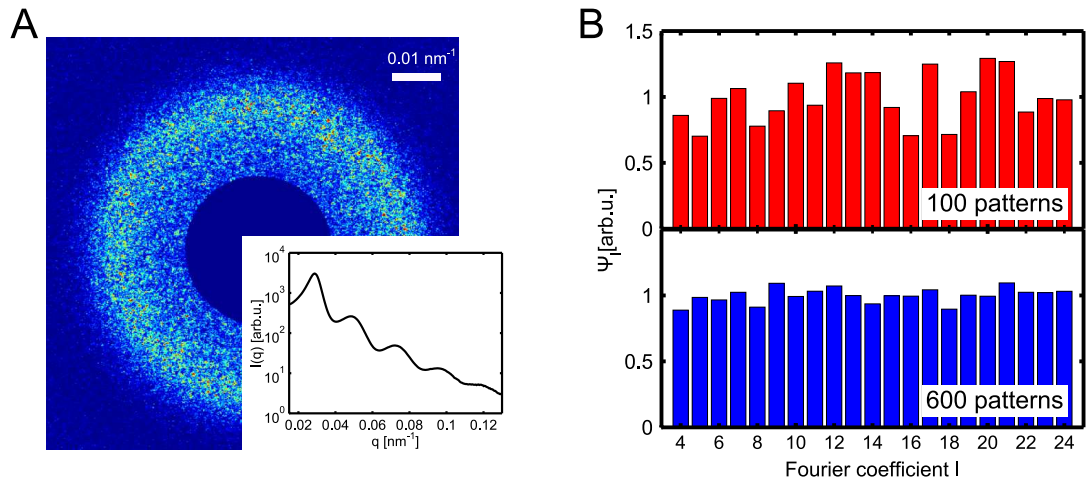}%
\caption{Experimental data. (A) Speckle pattern and $I(q)$ measured from a hard sphere colloidal glass. (B) $\Psi_l$ at $q=0.03$ nm$^{-1}$ averaging over 100 patterns (top) and 600 patterns (bottom)}%
\label{fig_exp}
\end{figure}

In order to test the cross correlation analysis discussed above, we performed a coherent scattering experiment on colloidal glasses (for more details see \cite{lehmkuehler2013}). The coherent small-angle x-ray scattering experiment has been performed at beamline P10 of PETRA III at DESY (Hamburg, Germany). The beamsize was defined by slits to 10 $\upmu$m$\times$ 10 $\upmu$m and the x-ray energy was chosen to 7.9 keV. A Princeton Instruments CCD detector (1300$\times$1340 pixels, pixel size 20 $\upmu$m) was placed at a distance of 5 m to the sample. Speckle patterns were measured from a colloidal hard sphere glass consisting of a 1:1 mixture of polymethylmethacrylat (PMMA) particles in decalin with radii of 125 nm and 84 nm, respectively. A sample scattering pattern is shown in Fig.~\ref{fig_exp} A together with the azimuthally integrated intensity $I(q)$ calculated from a pattern averaged over 1000 single frames. The grainy speckle structure is clearly visible in the pattern. In contrast to the simulation, the data were taken from a 3-dimensional sample (using capillary of 0.7 mm thickness).

In Fig.~\ref{fig_exp} B the results of the XCCA analysis for $q=0.03$ nm$^{-1}$ is shown for two different ensemble sizes. In contrast to the simulation data odd coefficients can be observed with amplitudes similar to the even ones. Furthermore, the results seem to be clearly depending on the statistics, i.e.~the number of patterns that were used for calculation of $\Psi_l$. Therefore, we need to study the statistical accuracy which is necessary for calculation of $\Psi_l$ and potential other factors that may lead to the appearance of odd XCCA coefficients in the simulation model.

\subsection{Statistical accuracy}
To estimate how many single speckle patterns have to be measured to detect a preferred order by XCCA from the viewpoint of simulations, the evolution of $\Psi_l$ with the number of arrangements can be studied. The results are presented in Fig.~\ref{fig:expconmerge} A for $l=6$ exemplary for four systems consisting of a different number of hexagons ranging from 4 to 1600. Here, $\Psi_6$ is calculated after each arrangement of hexagons and normalized to the final value, i.e.~$\Psi_6(2000)$ after 2000 runs. Similar results can be found for all other Fourier coefficients as well as other runs and were also reported recently for $C_l$ \cite{kurta}. Obviously, the number of hexagons studied does not influence the deviation of the final value of the orientational order. This is in agreement with recent simulations \cite{kirian}, where the signal to noise ratio was found to be independent from the number of particles. After approximately 50 speckle patterns $\Psi_6$ does not change significantly, after 200 patterns $\Psi_6$ stays inside a range of $\pm$ 10\%\ of the final value. To achieve statistics of less than 2\%\ at on the order of 1000 patterns have to measured in such two-dimensional systems. These observations agree to the experimental results in Fig.~\ref{fig_exp} B. Similar results are achieved for other Fourier coefficients $l$ and different sample systems, e.g.~the two component system consisting of pentagons and heptagons.

\subsection{Noise}
As discussed above, only even symmetries are allowed in the speckle patterns from electron densities with vanishing imaginary part. This is in particular true for the results of simulations, where odd symmetries are typically three to four orders of magnitude weaker than the even ones. The fact that they are not zero can be understood as shortcoming of the numerical accuracy of the calculation of the Fourier transform. In experiments the amplitude of odd symmetries is usually larger (see Fig.~\ref{fig_exp} B), partially caused by noise effects from the photon counting process. To estimate the influence of such noise, we chose arrangements of 400 hexagons.

The counting noise was modeled by random noise which was added to the calculated speckle patterns. Its magnitude is described by the intensity to noise ratio INR. For instance, an INR value of 10 means that the mean value of the added noise is ten times smaller than the pure signal without noise. Afterwards, the patterns were normalized to their mean values and $\Psi_l$ was calculated from the ensemble of patterns for a $q$-region that corresponds to the next-neighbor distance. To quantify the influence of the noise, the mean values $\langle\Psi\rangle_j$ were calculated for both odd and even symmetries, where $\langle\rangle_j$ denotes the average for either odd or even symmetries. The results are shown in Fig.~\ref{fig:expconmerge} B as function of the intensity to noise ratio INR.

At high INR values, the odd symmetries are almost zero. The computational accuracy of calculation of the Fourier coefficients is demonstrated by the finite and non zero value of the odd symmetries with increasing INR. Odd symmetries become observable for low INR with a simultaneous decrease of the magnitude of the even symmetries. In particular, for INR$\leq 5$ $\langle\Psi\rangle_\mathrm{even}$ starts to decrease while the odd coefficients increase. Finally, for INR$\leq 0.02$ the magnitudes of even and off symmetries are comparable. Remarkably, XCCA allows still at very low INR values around and below 0.1 the observation of dominant order, which is usually not possible using other methods.

\subsection{Non-planar wavefronts}
\begin{figure}%
\includegraphics[width=\columnwidth]{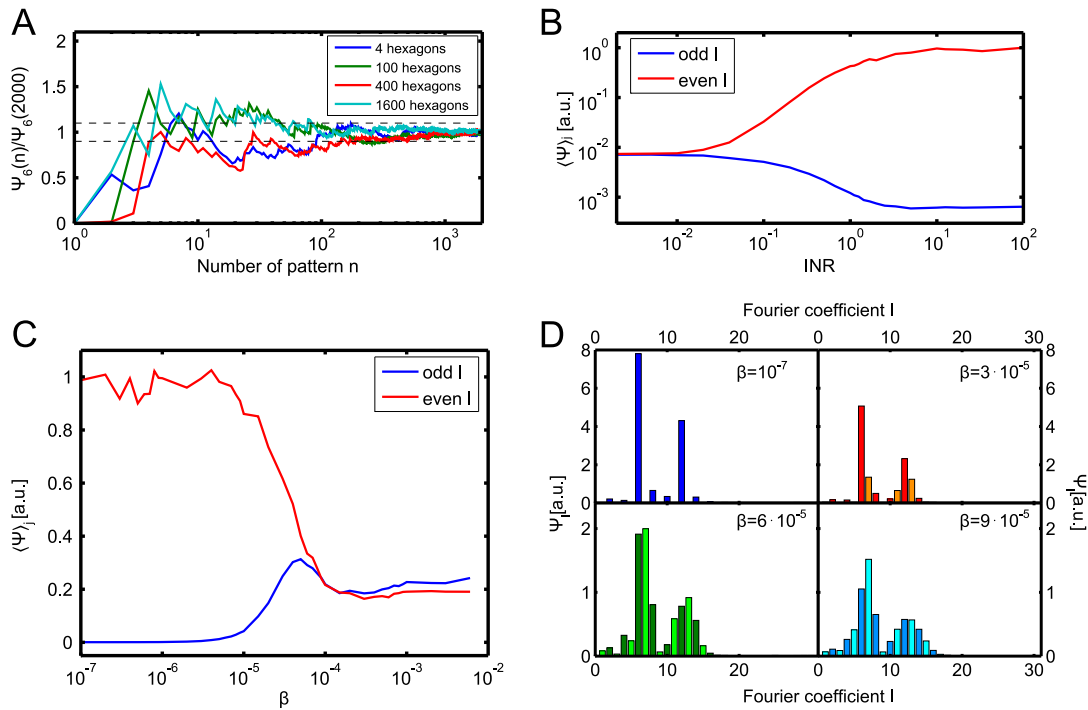}%
\caption{(A) Convergence of $\Psi_6$ (i.e.~$l=6$) for four hexagon systems. The dashed black lines represent a 10\%\ range of deviation from the final value. (B) Influence of detector noise on the even and odd symmetries in $\Psi_l$ as function of the intensity to noise ratio INR. The data are normalized to the mean of all even symmetries without noise (i.e.~$\mathrm{INR}\rightarrow\infty$). (C) Effect of a non-planar wavefront. Odd and even coefficients are shown as function of the dimensionless parameter $\alpha=\beta d^2$, the index $j$ denotes the average over either odd or even coefficients. The values are normalized to the mean of all even symmetries for a planar wavefront. (D) $\Psi_l$ for selected values of $\alpha$. The odd coefficients are highlighted by a lighter color. Data are normalized to the mean of all even symmetries at an planar wavefront.}%
\label{fig:expconmerge}%
\end{figure}

Beside the photon noise, odd coefficients can be caused by non-planar x-ray wavefronts \cite{rutishauser}. So far, our calculation were performed assuming ideal experimental setting, in particular a planar x-ray wavefront. Here we discuss the influence of a curved x-ray wavefront by replacing the modeled structure density $\rho({\bf r})$ with a so called Fresnel density \cite{sinha}
\begin{equation}
\rho_F({\bf r})=\rho({\bf r})\exp(i\alpha r^2),
\label{eq:fresdens}
\end{equation}
with ${\bf r}$ having its origin in the center of the computational box. Assuming no influence of beam defining aperture, $\alpha$ can be calculated via \cite{sinha}
\begin{equation}
\alpha=\frac{1}{2}\left(1+\frac{\Delta\lambda}{\lambda}\right)\cdot\frac{k_0}{L},
\label{eq:defalpha}
\end{equation}
with the monochromacity of the beam $\frac{\Delta\lambda}{\lambda}$, $k_0$ the absolute value of the incoming wave vector, and $L$ the distance between sample and detector. For convenience, we substitute $\alpha$ by the dimensionless parameter $\beta=\alpha d^2$, with $d$ denoting the particle size. A typical hard x-ray scattering experiment on colloidal particles in SAXS geometry ($\lambda=1.5$ \AA, sample-detector distance approx.~5 m, particle size $d= 100$--500 nm), results in typical values in the order of $10^{-6}<\beta<10^{-5}$. We studied the hexagon system at different values of $\beta$ as shown in Fig.~\ref{fig:expconmerge} C. At low $\beta$ values, no contribution of odd Fourier coefficients can be observed. This changes for $\beta\geq 10^{-5}$ where odd coefficients increase accompanied by a decrease of the even coefficients. For $\beta\geq 10^{-4}$ both are almost equal.

Fig.~\ref{fig:expconmerge} D shows $\Psi_l$ for some $\beta$ values. The increase of the magnitude of the odd coefficients is mainly limited to the contributions next to the dominant even coefficients, e.g.~$l=5,7$ at $\alpha=6\cdot 10^{-5}$. At larger $\beta$ values, other even components increase in addition, such as $l=8$. Nevertheless, the maxima around $l=6, 12$ are still observable. Thus, the shape of the x-ray wavefront may influence the results of an XCCA experiments significantly. This is of particular importance, if the wavefront is unknown. The influence of non-planar wavefronts is further investigated for cross-correlation studies with visible light, see \cite{schroerpre}.

\section{Summary}
In summary, by introducing a new correlation function $\Psi_l$ we demonstrated the feasibility of XCCA to detect the orientational order of a sample. We showed that additional Fourier coefficients of the cross correlation function $C$ can originate from the tiling of the computational box and are not an intrinsic feature of XCCA in dense systems.

We note that such a correlation function is also important for experimental situations where background and slit scattering are often present and need to be taken into account. Furthermore we show that a particular orientational order can already be detected if only few particles exhibit such an order. The study of orientational order as a function of the number of speckle patterns and the influence of random noise and non-planar wavefronts suggest that on the one hand at least some 100 single patterns have to be measured to detect a preferred orientational order with sufficient accuracy and on the other hand that noise and non-planar wavefronts influence the amplitudes of $\Psi_l$ significantly. The findings of this study will guide the way for application of the XCCA method in subsequent experiments on amorphous materials such as liquids and glasses. Naturally, further considerations have to be taken into account when three-dimensional systems are studied \cite{kurta2013} which we will discussed in subsequent theoretical and experimental work \cite{lehmkuehler2013,schroerpre}.

\ack{Acknowledgments}
We thank Birgit Fischer for sample preparation and support during the beamtime. Michael Sprung is acknowledged for experimental support. We like to acknowledge the Excellence Cluster "Frontiers in Quantum Photon Science" and the "Hamburg Center for Ultrafast Imaging" (CUI) for financial support.

\bibliography{bondordersim_jac}
\bibliographystyle{iucr}

\end{document}